# MULTI-LABEL SOUND EVENT RETRIEVAL USING A DEEP LEARNING-BASED SIAMESE STRUCTURE WITH A PAIRWISE PRESENCE MATRIX

*Jianyu Fan[1,2], Eric Nichols[2], Daniel Tompkins[2], Ana Elisa Méndez Méndez[2,3], Benjamin Elizalde[2,4], Philippe Pasquier[1]*

[1] Simon Fraser University, Surrey, BC, Canada
[2] Microsoft, Dynamics 365 AI Research, Redmond, WA, USA
[3] New York University, New York, NY, USA
[4] Carnegie Mellon University, Pittsburgh, PA, USA
Email：jianyuf@sfu.ca, epnichols@gmail.com, daniel.tompkins@microsoft.com, anaelisamendez@nyu.edu, bmartin1@cmu.edu, pasquier@sfu.ca

## ABSTRACT

Realistic recordings of soundscapes often have multiple sound events co-occurring, such as car horns, engine and human voices. Sound event retrieval is a type of content-based search aiming at finding audio samples, similar to an audio query based on their acoustic or semantic content. State of the art sound event retrieval models have focused on single-label audio recordings, with only one sound event occurring, rather than on multi-label audio recordings (i.e., multiple sound events occur in one recording). To address this latter problem, we propose different Deep Learning architectures with a Siamese-structure and a Pairwise Presence Matrix. The networks are trained and evaluated using the SONYC-UST dataset containing both single- and multi-label soundscape recordings. The performance results show the effectiveness of our proposed model.

## 1. INTRODUCTION

Humans have an inherent ability to match sound events based on acoustic similarity and the relationship between them [1]. Previous studies mainly focus on sound event detection (SED), investigating which sound events happen in an audio recording and when they occur [2]. In contrast, Sound event retrieval (SER) is retrieving audio recordings that are similar to a given input audio query [3, 4]. This similarity can be based on acoustic and/or semantic (symbolic) characterization [5]. SER has received far less attention than SED.

Due to a growing number of sound recordings, sound designers face numerous challenges. The tasks of searching and listening to audio recordings from a large database can be repetitive and time consuming. SER research enables developing audio retrieval systems for browsing sound and assisting sound design. With SER models, engineers can design recommendation systems for audio recordings. Sound designers will find a more streamlined workflow to add suitable sound effects for films and games.

Previous audio retrieval research mainly focuses on either acoustic similarity or categorization [4-8]. We neither simply use SED techniques to classify sound and retrieve the label, nor simply adopt audio fingerprinting to measure similarity. We consider both semantic similarity and acoustic similarity to be important for sound event retrieval. Audio recordings from different categories might sound similar, while recordings from the same category might sound different. This problem is more significant in the case of multiple labels (i.e., multiple sound events occur in one recording). Therefore, we propose a Siamese-structure to measure the similarity between pairs of audio recordings, which contain single-label or multi-label annotations. Since there is no established perceptual similarity measure between sound events, we designed a representation of an output matrix for the level of similarity between audio samples regarding multiple labels. This work focuses on the precision of the retrieval; efficiency is left for future work.

## 2. BACKGROUND

Audio fingerprinting computes a digital summary of an audio recording and matches against those stored in the database by comparing similarity [9]. The goal is to find the identical, similar, or distorted versions of a target recording. Audio fingerprinting has been used in music recommendation [6], video identification [10] and SED [11]. However, when searching for similar audio recordings based on sound events, we require both an acoustic similarity measure, and semantic similarity matching [3, 4].

To achieve this, we propose using a Siamese Neural Network (SNN), which contains two identical sub-networks that share weights [12]. SNN have been explored for representations of and content understanding in text, voice, images and video [12-15]. Regarding sound events, Jiménez et al. proposed an ontology-based neural network for SED, which considers a structured relationship between class labels [16]. The authors use an SNN to compute embeddings to preserve the ontology information. Manocha et al. designed a system performing single-label sound event retrieval. They propose an SNN approach to obtain embeddings of single-labeled audio recordings (i.e., only one sound event exists in a recording) [4]. Then, the authors use K-Nearest Neighbors of embeddings as the retrieved results. Their evaluations indicate that the model captures semantic similarity between recordings. Inspired by the previous study, we designed an SNN for multi- and single-label SER. To the best of our knowledge, this is the first work studying multi-label SER based on deep-learning approaches.

## 3. APPROACH

### 3.1. Architecture of the Proposed Deep Learning Siamese-structure with a Pairwise Presence Matrix

We propose an SNN-based approach to measure the level of similarity between a pair of audio samples. Fig. 1 shows our proposed model. First, we compute the log-Mel spectrograms of multiple frames of audio recordings. Next, these raw features are passed to the VGGish model [17], a SED model trained on a large-scale audio dataset, to obtain latent embedding vectors (See 5.1 for details). Then, a pair of embedding vectors is passed to a Siamese-structure.

An SNN typically consists of two twin networks [12]. Each sub-network takes one input. Typically, to train an SNN, researchers use the contrastive loss function to measure the distance between a pair of input samples. If inputs are similar, then it should predict 1, otherwise 0. However, because our audio samples contain multiple labels, instead of using a contrastive loss, we concatenate the output embeddings from the Siamese-structure and pass it to a multilayer perceptron (MLP) network.

The Siamese-structure consists of 3 layers. Each layer is composed of 128 neurons. The outputs of the Siamese-structure are concatenated into a vector, which is passed to an MLP with 2 layers. The first layer consists of 256 neurons and the second layer contains 128 neurons. After that, we adopt the attention module proposed by Kong and Yu [18, 19]. As shown in (1), the predicted results can be weighted on information from different time steps, and ignore irrelevant sound segments such as silences. $v(\cdot)$ is a Softmax function that normalizes along time steps to determine how much an embedded feature $h_t$ should be attended to or ignored. $f(\cdot)$ are the prediction results of the output layer, which denote the classification output of a $h_t$. The ReLU non-linearity activation function is applied to all layers. We also added batch normalization in each layer and applied a dropout of 0.5 between all layers during training.

$$y(h) = \frac{1}{\sum_{t=1}^{T} v(h_t)} \sum_{t=1}^{T} v(h_t) f(h_t) \qquad (1)$$

Regarding the output of the proposed model, we define a three-column *Pairwise Presence Matrix* to represent the situations of each category for a pair of audio inputs so as to measure the level of similarity between them. For each category, we use one-hot encoding to represent three situations: both inputs contain it, neither contains it, or one of the inputs contains it. The level of similarity is the exact number of present and absent categories that two audio samples have in common. For example, the level of similarity of two samples in Fig. 2 is 4. For the output layer, we use a Softmax function with a categorical cross-entropy loss function.

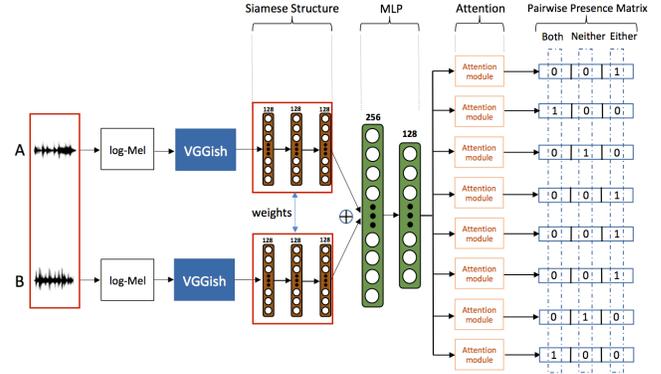

**Fig. 1**. Proposed Deep Learning Siamese-structure with a Pairwise Presence Matrix

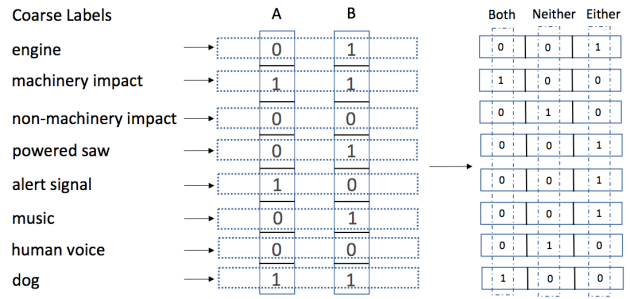

**Fig. 2**. Pairwise Presence Matrix

Given an input audio query, we first create audio pairs between the query and samples in the database (i.e., training set). Then, we pass the pairs to the network to obtain the presence matrix. By summing the first and the second column of the matrix, we obtain the level of similarity (i.e., exact number of present and absent categories that two audio samples have in common) between the audio query and samples in the database. All samples in the database are ranked according to their similarity to the query and then the top K samples are returned.

We believe both sematic categorical information and acoustic similarity are important for multi- and single-label SER. By using the VGGish model [17], and defining a pairwise presence matrix, we incorporate the semantic categorical information for the SER task. While comparing the embeddings of two audio samples together, we include the information of acoustic similarity.

## 4. DATASET

### 4.1. SONYC-UST dataset

We used the SONYC-UST dataset [20] for the audio retrieval task. SONYC-UST is a dataset extracted from the Sounds of New York City project [21] for mitigating urban noise pollution. Researchers collected recordings from 50 different sensors deployed in New York City. The dataset contains 2794 samples in total. There are 2351 samples in the training set and 426 samples in the test set, which contains 105 multi-label and 311 single-label recordings. All recordings are 10 seconds long and were recorded with microphones at identical gain settings. Researchers recruited individuals to provide weak labels for recordings based on a taxonomy involving 8 coarse- and 23 fine-grained categories [20]. We used coarse categories for the retrieval task (Fig. 2). This is challenging because various heterogeneous sources of noise pollution may overlap within the same acoustic scene and some categories, such as engine and machinery impact, sound similar to each other.

### 4.2. Creating Pairs for Training

Based on the weak labels, we create pairs of inputs for our model. When creating pairs, we intend to have balanced pairs regarding the coexistence of each category between a pair of samples. For each category, we selected a target sample (i.e., the first sample from the training set) and created 60 pairs by selecting: 1) 30 samples from the training set that have the same situation as the target sample (i.e., both samples contain this category, or neither contain this category) and 2) 30 samples from the training set that are opposite to the target sample (i.e., one of the samples contains this category). Then, we repeat the process by selecting the second sample as the target sample, and continue until the last sample is used as the target sample. After iterating through all the eight categories, we removed duplicate pairs and ended up having 218173 pairs in the training set. Since our goal is to perform retrieval, during the testing stage we create audio pairs consisting of one item from testing set (i.e., a query sample) and another item from the training set (i.e., the database). The weak labels of a pair of audio samples are converted to the pairwise presence matrix as described in Section 3.

## 5. EXPERIMENTAL SETUP

### 5.1. Feature Extraction

Given one-second audio excerpt, the VGGish model [17] computes the log-Mel spectrograms as raw features and then generates a 128-D embedding vector. Regarding the SONYC-UST, since each recording is 10 seconds, the dimensionality of the output of VGGish is 10×128. We also evaluated the performance of our model when only using log-Mel spectrogram. We converted the sampling rate to 22050 HZ, and chose the window size of 4096 and a step size of 2048. We ended up having a 108×128 feature vector for each audio recording.

### 5.2. Variations of Proposed Network: Single-model and Multi-model with or without Siamese-structure

We explored three more variations of our proposed network: 1) a version without the Siamese-structure, in order to evaluate the contribution of the Siamese-structure. 2) We split our original single-model architecture into eight sub-models, each corresponding to one of the eight coarse categories. 3) We evaluated multiple sub-models without Siamese-structure.

### 5.3. Network Settings

We trained each model using the RMSProp optimizer with a batch size of 128 samples, a learning rate of $1\times10^{-3}$. Due to the limited size of the dataset, we first select 10% of the training set as a validation set to determine the best number of epochs based on the validation error. Then, we combine the training set and the validation set together to train the model with the previously determined number of epochs.

### 5.4. Evaluation Metric

We use mean Average Precision ($mAP$) as the metric for the retrieval task. The $mAP$ can quantify the precision of a sorted retrieved list with the number of true positives leading the list. Since we have multiple labels, we adopted a $mAP_s@K$ metric [22], which is defined as follows:

$$Num\ of\ positive\ hits = \sum_{i=1}^{K} \Pi(r_{i,q} \geq s) \quad (2)$$

$$AP_s@K = \frac{1}{Num\ of\ positive\ hits} \sum_{n=1}^{K} \left(\frac{\Pi(r_{n,q} \geq s)}{n} \sum_{i=1}^{n} \Pi(r_{i,q} \geq s)\right) \quad (3)$$

$$mAP_s@K = \frac{1}{Q}\sum_{q=1}^{Q} AP_s@K \quad (4)$$

where $Q$ denotes the length of query set (i.e., test set). $K$ denotes the Top $K$ retrieved items, $s$ denotes the similarity threshold. To consider a retrieved sample as a positive hit, the level of similarity has to be equal or greater than a threshold ($s$). For example, if the threshold is 7, the level of similarity has to be 7 or 8 between the retrieved sample and the query so that the retrieved sample can be considered a positive hit. $r_{i,q}$ denotes the number of identical labels (out of 8) between the q-th query recording and the i-th database recording, $\Pi(\cdot) \in \{0, 1\}$ is an indicator function. $mAP_s@K$ ranges between 0 to 1. A high value means that more top instances in the retrieved list are considered as positive hits.

## 6. PERFORMANCE ANALYSIS

Because there are no other similar approaches in the literature, our baseline consists of randomly retrieving the top K recordings for a given query. All our architectures significantly outperformed the baseline as seen in Fig. 3. We show the performance obtained when the query was a single- or multi-label recording. We adjusted the similarity thresholds from 8 to 7 and selected different values of $K$, 1, 5, 10, 30, 50, and 100.

From Fig. 3, we observed that having the Siamese-structure always performs better than without it, which indicates its effectiveness to preserve similarity. Moreover, we observe that the $mAP_8@K$ of multi-label SER is the lowest among all experiments, implying that perfectly matching all categories between two samples is challenging when there are multiple sound events. Note that $mAP_8@K$ of single-label SER is fairly high. This is because distinguishing single sound event is much easier. When decreasing the similarity threshold from 8 to 7, the task is easier and performance improves significantly. Even with allowing only 7 identical labels (out of 8), the retrieved sample is still satisfying in most cases.

We evaluated the contribution of the attention module. We found that attention is better in the case of $mAP_8@K$ for single-labeled SER, $mAP_7@K$ for multi-labeled SER and for most $K$ values of $mAP_7@K$ for single-labeled SER. Fig. 3 shows that in the case of $mAP_8@K$ for multi-label SER, the architecture without attention works better.

Note that the multi-model architectures perform better than the single-model architectures (Fig. 1) in $mAP_8@K$ for multi-labeled recordings. Though multiple sub-models can learn specific semantic information for each category, the single-model incorporates well the inter-class differences of the coarse categories. Regarding features, we found that VGGish features perform significantly better than the log-Mel spectrogram features in all tests. This is because VGGish embeddings come from a pre-trained model, as opposed to log-Mel features.

## 7. CONCLUSION

This work investigates the both single- and multi-label SER problem. We present a Siamese-structured network to incorporate both acoustic similarity and semantic similarity information. Moreover, we design a pairwise presence matrix for the representation of the level of similarity between multiple labels. We compared the performance of our model with several variations, and our results show the effectiveness of the model for the retrieval task focusing on the precision of the retrieval task. More information can be found at http://metacreation.net/multi-label-sound-event-retrieval/. For future works, we will enhance the computational efficiency of the retrieval pipeline. We will also work on building larger datasets in order to scale these models to more categories.

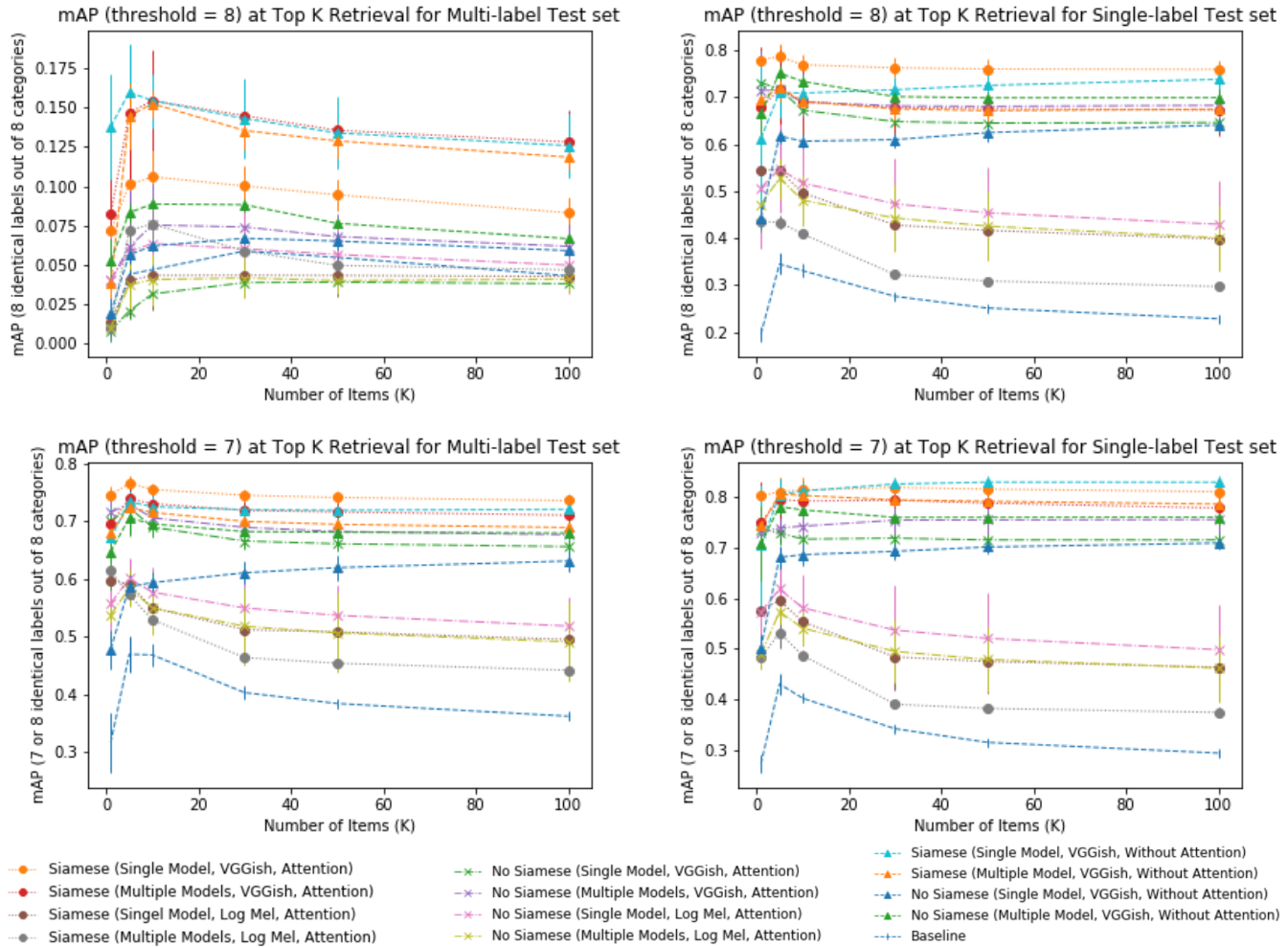

**Fig. 3**. Performance based on $mAP_s@K$ for the baseline, the Single-model and Multi-model with or without Siamese-structure